30

# PHASE BEHAVIOR AND EMULSION STABILITY OF THE AOT/DECANE/WATER/NaCl SYSTEM AT VERY LOW VOLUME FRACTIONS OF OIL

Yithanllili Bastidas[1], Lisset Hernández[2], Issarly Rivas[3], Kareem Rahn-Chique[3], German Urbina-Villalba[3*]

[1]Universidad Nacional Politécnica de la Fuerza Armada (UNEFA), Núcleo Carabobo, Naguanagua, Venezuela.
[2]Instituto Universitario Santiago Mariño, Sede Maracaibo, Venezuela.
[3]Instituto Venezolano de Investigaciones Científicas (IVIC), Centro de Estudios Interdisciplinarios de la Física, Apartado 20632, Edo. Miranda, Venezuela.

**Abstract**   The stability of a ternary system composed of decane/water/Aerosol-OT and salt is revisited. Phase diagrams and emulsions similar in composition to those previously studied by Hofman and Stein [Hofman, 1991] were made. According to our results, and contrary to the common experience, these systems exhibit a maximum of stability very close to the balance zone.

## 1. INTRODUCTION

The phase behavior of ternary systems composed of oil, water and surfactant markedly depends on the relative affinity of the amphiphile for the immiscible liquids. This chemical compatibility relies on its hydrophilic and hydrophobic moieties. The spatial separation of these two regions along its molecular structure renders the dual empathy of the surfactants for both liquids, along with a pronounced tendency to adsorb to oil/water interfaces.

Very high surfactant concentrations induce the homogeneous mixture of oil and water into one thermodynamically stable phase. Lower surfactant concentrations guarantee the endurance of the original phases, with substantial changes in their internal structure. If the surfactant is mostly soluble in water, it promotes the formation of an aqueous solution of oil-swollen micelles (o/w microemulsion) in equilibrium with an exceeding oil phase. A triangular phase diagram which illustrates this scenario is known as a Winsor I diagram. Conversely, a Winsor II diagram shows a two-phase region containing a w/o microemulsion in equilibrium with an exceeding water phase. Since swollen micelles form spontaneously, the radii of curvature of these structures is regarded as the spontaneous curvature of the surfactant. It is assumed that any radius of a drop different from the radius of curvature of the surfactant implies an additional excess energy which destabilizes the system [Langevin, 1992; De Gennes, 1982; Kellay, 1994; Binks, 1989; Safran, 1986; Cates, 1988]. This is one of the reasons why emulsions can only be kinetically stable and not thermodynamically stable as their microemulsion analogs.

When the affinity of the surfactant for oil and water is similar, the mixture of oil, water and surfactant generates a three-phase region in the phase diagram (Winsor III diagram) [Shinoda, 1967; Kunieda, 1981]. Two of these phases are basically composed of water and oil, and the third phase essentially contains all the surfactant in the system. This additional phase is usually observed in the middle of the reservoir. Depending on its physical properties, it is classified as either a liquid crystal (lamellar, cubic, hexagonal, etc), or a bi-continuous microemulsion (a phase in which continuous regions of oil and water coexist).

It is possible to induce a Winsor I – Winsor III - Winsor II transition (and vice-versa) changing the relative affinity of the surfactant for the oil and water phases [Salager, 1982]. This can be done increasing the ionic strength of the water phase, changing the composition of the oil, decreasing the temperature of the system, etc. During these changes it is observed that the volumes of the oil and water phases vary monotonically and in the opposite direction (one increases while the other decreases). Hence, it is possible to equalize these two volumes changing a formulation variable systematically. In this case, it is said that the surfactant is at its balance point, where its affinity for oil and water is perfectly balanced.

The phase behavior described above is extremely interesting in its own right. Especially if one realizes that the interaction between the different constituent molecules of the system is always attractive [Urbina-Villalba, 1995, 1996]. Hence, in the case of two neat phases like oil and water, it is the distinct magnitude of the intermolecular forces (the cohesive energy of the liquids) which disfavor their mixture and drives their separation. In microscopic terms: the type and number of spatial configurations by which the electronic density of the molecules fits the void spaces without overlapping. This description is closely related to the actual density of the liquids in the system. Hence, it does not appear to be a coincidence that the third phase of a Winsor III system, generally appears in the middle of the oil and water phases. In the case of more complex systems (like microemulsions) in which the extent of the interface is considerable, it appears to be the free energy of the interface who determines the phase behavior of the system.

If a specific composition corresponding to a two-phase region of a ternary system is stirred vigorously, the formation of a mixed dispersion of drops of oil-in-water and water-in-oil is provoked. After a finite time, it is observed that a system originally corresponding to a Winsor I diagram produces a "direct" oil-in-water (o/w) emulsion, while a system starting from a Winsor II diagram creates an "inverse" water-in-oil (w/o) emulsion. In each case, the type of emulsion is the result of the prevailing coalescence frequency between the drops of water and oil. Moreover, the stability of the resulting emulsion increases with the increasing surfactant concentration. This is attributed to the effectiveness of the surfactant in generating repulsive barriers between the drops. The kinetic stability of these systems generally conforms to the Derjaguin-Landau-Verwey-Overbeek (DLVO) theory [Derjaguin, 1941; Verwey, 1948].

Curiously, it is observed that when a ternary Winsor III system approaches its balance point, the interfacial tension between the middle phase and the other two phases attains ultralow values [Aveyard, 1986, 1987, 1988; Binks, 1993, 1999; Salager, 1982]. This property is commonly used to synthesize emulsions of nanometric particle size [Solé, 2002]. According to the Gibbs adsorption isotherm, minimum values of the tension indicate maximum surfactant adsorption. Besides, it is unlikely that the physical properties of the exceeding oil and water phase equalize those of the middle phase at the balance point. Therefore it is astonishing that the emulsions prepared by vigorous stirring of a ternary system at the balanced state leads to very unstable emulsions [Shinoda, 1967; Bourrel, 1979; Salager, 1982].

Several hypotheses had been forwarded to explain this instability, but up to our knowledge, only one has been successfully proven. It has been argued that:

1) The surfactant is more stable in the middle phase, and tends to aggregate in it near the balance state. Hence, any oil/water interface formed during stirring is depleted from these molecules.
2) The surfactant has the same tendency to go to the interface than to the other two bulk phases. Therefore, it does not preferentially adsorb to the interface at the balance point.
3) The drops formed in the balance state are highly deformable [Danov, 1993], and as a consequence, most likely to coalesce than non-deformable drops.

Each of these hypotheses has several drawbacks. First, it is important to recognize that when the mixing of a balanced system is stopped, the initial state does not correspond to either a water-in-oil or an oil-in-water emulsion. Due to the existence of the additional third phase, these systems are not equivalent to conventional emulsions, and therefore, any perturbation of their apparent homogeneity after stirring is regarded as a sign of instability. Second, hypothesis 1 does not explain why a dispersion of drops of the third phase is not formed (and preserved) within the excess oil and water phases. Third, hypothesis 2 is inconsistent with the fact in the balance state, almost all the surfactant population concentrates in the middle phase. Fourth, it is unclear whether or not deformable drops are more unstable than deformable drops [Toro-Mendoza, 2010].

Probably the most successful theory of stability in the balance state is the one introduced by Kabalnov, Wennerström and Weers [Kabalnov, 1996a, 1996b]. According to these authors, the process of coalescence in the balance state implies the formation of a hole between deformed droplets. The free energy of hole formation depends markedly on the bending elasticity of the surfactant (K) and its spontaneous curvature. The spontaneous curvature strongly affects the free energy penalty for nucleation of a critical hole in an emulsion film. This dependence comes from the fact that the surfactant monolayer is strongly curved at the edge of a nucleation pore.

The above findings were tested using a system composed of octane, water and $C_{12}E_5$ [Kabalnov, 1996b]. It is well known that the solubility of ionic surfactants in the water phase can be decreased increasing the temperature of the system. Thus, a Winsor I – Winsor III – Winsor II transition is observed T = 33 °C. Near the balance point, the stirring of the system produces macroemulsions of the oil and water phases emulsified in each other, while the middle phase does not co-emulsify with them, and separates within the first hour after emulsification. Within 0.2 degrees on both sides of the balance point, macroemulsions are very unstable. However, within the following 0.15 degrees a spectacular increase of stability of three orders of magnitude occur on either side of the phase inversion temperature.

Back in 1991, Hofman and Stein [1991] studied the temporal evolution of a (decane + $CCl_4$) /water emulsion stabilized with either Aerosol-OT (AOT) or sodium oleate (NaOl). Gemini-surfactants are the only anionic surfactants able to reach ultralow interfacial tensions in the absence of co-surfactants or salt. Instead, NaOl exhibits a minimum constant value of 0.1 mNm at interfacial saturation. On this basis it was expected, that the drops of the emulsion will only show a substantial degree of deformation in the presence of AOT. Using a Coulter Counter the authors studied the variation of the average radius (R(t=0) = 1.25 μm) of the emulsion as a function of time, and used this information to compute the temporal variation of the number of drops. The procedure allowed studying the change of the mixed flocculation/coalescence rate as a function of the salt concentration. As a result they concluded that AOT stabilized emulsions are less stable than NaOl-stabilized ones *in the field of slow coagulation*. The data was insufficient to establish a conclusion at high ionic strength, but the results suggest that in that case, AOT-stabilized emulsions could be more stable. Recently we published theoretical simulations which resemble the experimental conditions of Hofman and Stein, finding that the experimental variation of the flocculation rate can only be accurately reproduced assuming deformable drops for both systems. Otherwise there are surfactant concentrations of AOT and NaOl for which the drops of the emulsions behave as non deformable particles, and some other concentrations for which they behave as deformable drops.

Barnes and Prestidge studied the temporal evolution of the turbidity of polydimethylsiloxane dispersions in water [Barnes, 2000]. The deformability of the drops was changed by addition of triethoxymethylsilane to the polymer. They concluded that deformable droplets are more stable than equivalent non-deformable drops.

Due to its possible application in secondary oil recovery, the physicochemical properties of AOT has been extensively studied [Nave, 2000a, 200b, 2002], although no relevant peculiarities of its molecular structure were found. Binks *et al*. [Binks, 1999] studied the evolution of o/w and w/o silicon macroemulsions with respect to creaming and coalescence. For o/w emulsions the stability of the dispersions markedly decreases when approaching the Winsor I/Winsor III boundary. Instead, for w/o the stability increases close to the boundary where three phases form. Three phase emulsions were found to be extremely unstable.

It is clear from above that in order to advance in the comprehension of the stability problem in the balance zone, a study of the flocculation-coalescence rate of oil-in-water nanoemulsions could be very valuable. These studies can now be done using a new procedure developed by our group [Rahn-Chique, 2012]. As shown in the study of Toro-Mendoza [2010] the deformation of emulsions drops changes as a function of size. Very small (sub-micron) drops are unlikely to deform. Hence, these studies can evaluate the relationship between deformation and stability in the balance zone. If the variation of the average radius is followed as a function of time and the type of oil is varied, the contribution of Ostwald ripening to the destabilization of the system can be appraised. The proposed research starts from the localization of the balance zone. This region depends on the composition of the system (salinity, volume fraction of oil, etc). In this preliminary report, the same system employed by Hofman and Stein is revisited. First, the variation of the balance zone with the volume fraction of oil is studied. Then, the rate of creaming of decane-in-water macroemulsions is monitored by means of a Quickscan (Coulter). Unlike previous studies, the present measurements indicate the existence of an unexpected maximum of stability very near the balance zone.

**2. EXPERIMENTAL DETAILS**

2.1 Materials

Sodium Docusate (AOT) $C_{20}H_{37}NaO_7S$ (FW 444.55 g/mol, mp 173-179 C, d = 1.1 g/cm$^3$, D1685, CAS 577-11-7) was obtained from Sigma. Sodium chloride NaCl (pro-analysi, 99.5%), and Decane $C_{10}H_{22}$ (for synthesis,

FW 142.28 g/mol, d= 0.728 – 0.732 g/cm$^3$) were obtained from Merck and used as received. Millipore's distilled and deionized water (1.1 μS/cm at 25°C) was used in all experiments.

## 2.2 Phase Diagrams

A mother solution of AOT was prepared. The solid was previously dissolved in a beaker under magnetic stirring, and then transfused to a calibrated flask. One vial was used for each specific composition of the system. The AOT-solution, decane, and salt were added in that order. The top of each vial was covered with plastic paper and then tightly closed. The systems were gently mixed during one hour using a Varimix (Thermo), and then let to settle for one week or until equilibrium was attained.

The number of liquid phases in the systems was detected by direct observation (Figure 1). In binary systems, a laser beam was used to evidence the presence of swollen micelles. The passage of a laser beam through a homogeneous liquid cannot be observed, although the beam can be detected at the other side of the reservoir. Hence, the beam appears to be "cut" by the liquid. A microemulsion phase can be identified because it contains swollen micelles that scatter the laser light, allowing the passage of the beam to be observed within the liquid.

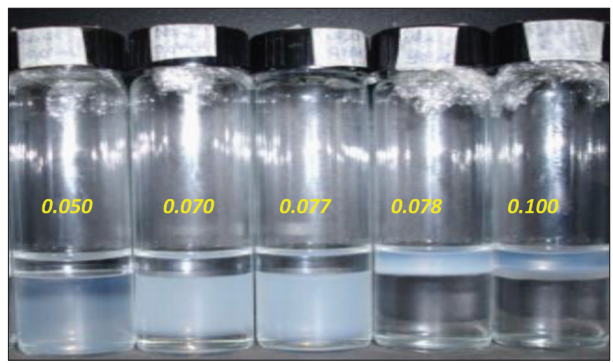

Figure 1: Phase behavior of the decane/water/AOT system as a function of the NaCl concentration (yellow labels (M)) for a surfactant concentration of $C_s$ = 1.0 x 10$^{-2}$ M, and a volume fraction of oil of $\Phi$ = 0.2. The location of the surfactant during the Winsor I – Winsor III – Winsor II transition is evidenced by the turbidity of the surfactant-rich phase.

Liquid crystal phases were detected using two polarizers, one on each side of the vial, initially disposed at an angle of 90° degrees. The light that enters through one side of the vial is polarized, but its angle of polarization is changed due to the presence of the liquid crystal. Hence, this structure is assumed to be present if a brightness can be detected despite the initial position of the polarizers. In the case of bicontinuous phases, an intermittent glow is only observed when the system is gently shaken.

A more strict classification of the homogeneous phases (into $L_\alpha$, cubic, hexagonal, etc) was not endeavored.

The compositions selected in this study attempted to recreate as possible those of the experiments of Hofman and Stein. For their AOT emulsions these authors used a volume fraction of oil $\Phi$ = 0.0083, a surfactant concentration (Cs) of 4.2 x 10$^{-4}$ M, and salt concentrations between 0.05 and 0.5 M. According to their tension measurements, and for Cs = 4.2 x 10$^{-4}$ M, the balance zone occurs at 0.05 M $C_{NaCl}$ ($\gamma$ < 10$^{-3}$ mNm).

Here, phase diagrams were built at $C_s$ = 1.0 x 10$^{-2}$, 2.2 x 10$^{-3}$, and 4.2 x 10$^{-4}$ M. For this purpose, volume fractions of oil equal to 0.001, 0.01, 0.05, 0.1, 0.2 and 0.3 were used. The salt concentration mostly varied between 0.01 and 0.1 M.

## 2.3 Emulsions stability with respect to creaming

Emulsions were prepared with an Ultraturrax at a velocity of 17,500 r.p.m. They were allowed to rest during 1 hour to eliminate the foam. The average size of the drops varied between 0.7 and 2.3 μm depending on the composition. The polydispersity was high (C.V. = 30% – 40%) and in several cases bimodal drop size distributions were observed. In order to identify the type of emulsion occurring at each composition, the conductivity of the dispersions was measured as a function of the ionic strength, for the same values of $C_s$ and $\Phi$ employed in the construction of the phase diagrams.

The stability of the emulsions with respect to creaming was monitored using a Quickscan apparatus (Coulter). Originally the emulsions were turbid, and progressively clarify from the bottom of the recipient to the top. The time required for each system to attain a Transmittance of 50% at half the height of the tube was used as a stability criteria. For this study, four sets of experiments were made:

Set I: At fixed values of $C_{NaCl}$ = 0.01 M and $\Phi$ = 0.0083, the surfactant concentration was varied (1.0 x 10$^{-5}$ M ≤ $C_s$ ≤ 9.5 x 10$^{-2}$ M).

Set II: At a fixed value of $C_{NaCl}$ = 0.05 ($\Phi$ = 0.0083) the surfactant concentration was varied between 10$^{-7}$ M and 5 x

Yithanllili Bastidas, Lisset Hernández, Issarly Rivas, Kareem Rahn-Chique and German Urbina-Villalba

$10^{-4}$ M ($C_s$ = 1.0 x $10^{-7}$, 5.0 x $10^{-6}$ M, 1.0 x $10^{-5}$, 1.0 x $10^{-4}$, 1.5 x $10^{-4}$, 3.5 x $10^{-4}$, 4.3 x $10^{-4}$, 5.0 x $10^{-4}$).

Set III: At fixed values of $C_s$ = 2.2 x $10^{-3}$ M and $\Phi$ = 0.05, the effect of the ionic strength was studied at: 0, 0.01, 0.05, 0.06, 0.07, 0.09, and 0.1 M.

Set IV: At $C_s$ = 4.3 x $10^{-4}$ M and $\Phi$ = 0.0083, the salt concentration was varied between 0.01 and 0.25 M ($C_{NaCl}$ = 0, 0.010, 0.020, 0.030, 0.040, 0.050, 0.075, 0.100, 0.125, 0.150, 0.175, 0.200, 0.225, 0.250 M).

The first two sets explore the behavior of the decane / AOT / water system with respect to variation of the surfactant concentration. The last two sets show the effect of the ionic strength.

## 3. RESULTS AND DISCUSSION

The phase diagrams evaluated in this work are shown in Figures 2-4. The decane/AOT/water mixture shows one, two or three phases depending on the amount of surfactant and the salinity of the aqueous phase. The transition Winsor I – Winsor III – Winsor II is observed with the two higher concentrations tested ($C_s$ = $10^{-2}$ and 2.2 x $10^{-3}$ M). In Fig. 2 this transition occurs through a bicontinuous microemulsion, while in Fig. 3, it happens through a liquid crystal phase. Low salt concentrations induce the dissolution of the surfactant in the aqueous phase ($\underline{2}$ system). As the ionic strength increases the surfactant is salted out from the aqueous phase, generating a surfactant-rich middle phase. Further addition of salt induces the transference of the amphiphile to the oil phase in the form of inverse water-swollen micelles ($\overline{2}$ system).

The minimum salt concentration necessary for the attainment of a three-phase system changes with the surfactant concentration and the volume fraction of oil. However, in most cases it appears between 0.06 M < $C_{NaCl}$ < 0.09 M.

The behavior of the ternary system at $C_s$ = 4.3 x $10^{-4}$ M is different from the rest. The amount of surfactant appears to be too low to generate a middle phase. The system apparently changes directly from Winsor I to Winsor II ($\underline{2}$ → $\overline{2}$). The transition occurs between the concentrations of 0.08 and 0.10 M of NaCl. However, it is difficult to determine the preferential location of the surfactant at this small concentration by the use of a laser. Only a weak discontinuous beam is barely appreciated within the microemulsion phases. Moreover, it is not possible to ascertain that a very thin middle phase does not occur within a thin region of salinity not tested.

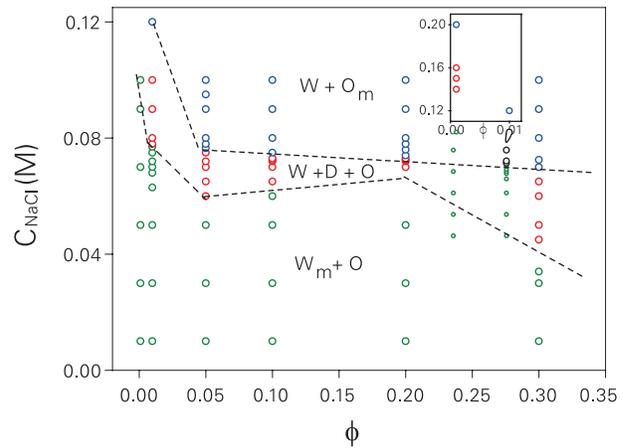

Figure 2: Phase diagram of the ternary system for a surfactant concentration $C_s$ = 1.0 x $10^{-2}$ M.

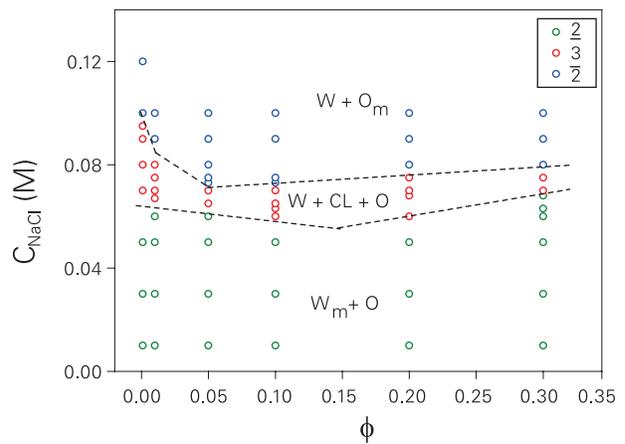

Figure 3: Phase diagram of the ternary system for a surfactant concentration $C_s$ = 2.2 x $10^{-3}$ M.

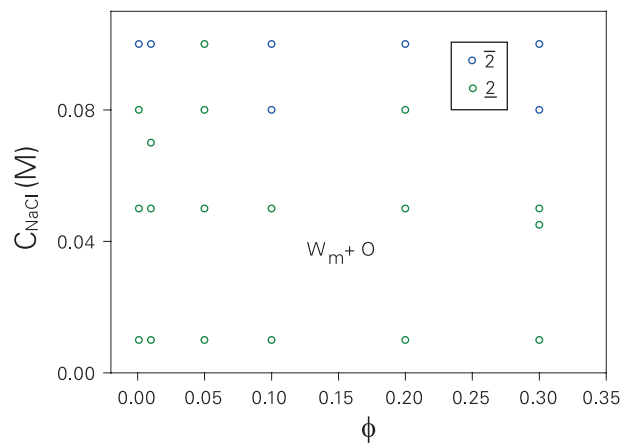

Figure 4: Phase diagram of the ternary system for a surfactant concentration $C_s$ = 4.2 x $10^{-4}$ M.

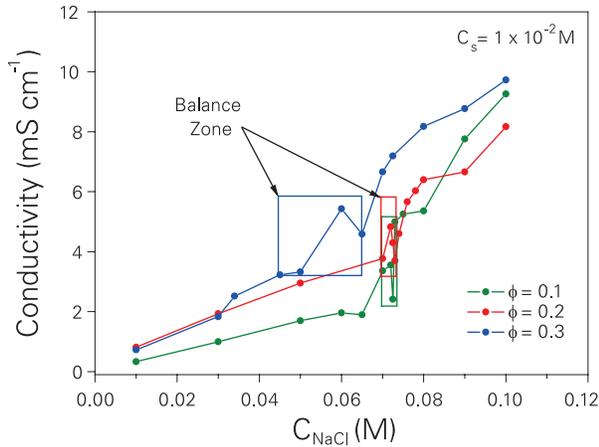

Figure 5: Conductivity of the systems with $C_s = 1.0 \times 10^{-2}$ M, and $\Phi = 0.1, 0.2$ and $0.3$.

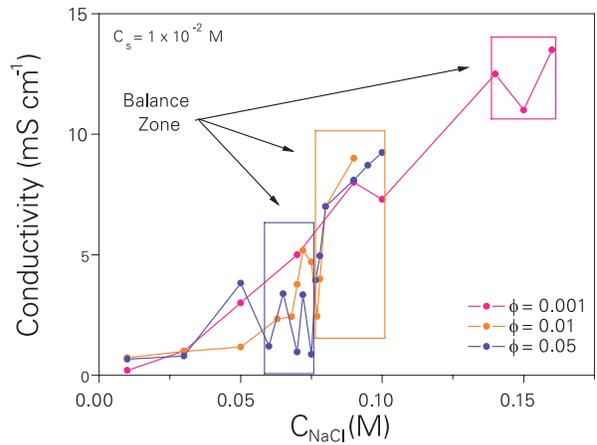

Figure 6: Conductivity of the systems with $C_s = 1.0 \times 10^{-2}$ M, and $\Phi = 0.001, 0.01$ and $0.05$.

Figures 5-9 show the conductivity of the emulsions resulting from the vigorous stirring of the systems whose equilibrium composition was previously studied. Except for an erratic behavior occurring near or within the three-phase region, the conductivity increases monotonically with the salt concentration indicating the occurrence of oil-in-water emulsions. In fact, a linear dependence of the conductivity with $C_{NaCl}$ is observed for $C_s = 4.3 \times 10^{-4}$ M. The behavior of this parameter within the balance zone can be attributed to the lack of stability of the emulsions formed (see below). Notice how the balance zone moves with the volume fraction of oil and the surfactant concentration. In most cases it differs considerably from the value of $C_{NaCl} = 0.05$ M which corresponds to the salinity where the minimum value of the surface tension is observed in macroscopic decane/water systems [Hofman, 1991].

Figure 10 shows the dependence of the stability of the ternary systems as a function of the surfactant concentration at $\Phi = 0.0083$, $C_{NaCl} = 0.01$ M. On the basis of DLVO theory, the stability of an oil-in-water emulsion is expected to increase with the surfactant concentration, and reach a plateau. This will reflect the progressive increase of the surfactant surface excess as a function of the total surfactant concentration, as well as the raise of the charge of the drops up to the saturation of their interface. However, in the case of an ionic surfactant, the increase of the surfactant concentration also augments the ionic strength of the aqueous solution. This effect screens the charge of the

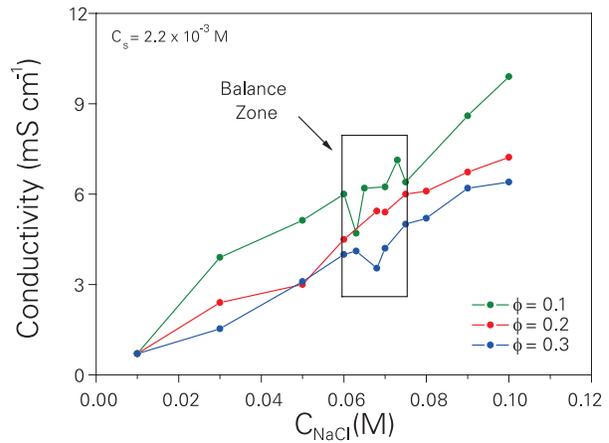

Figure 7: Conductivity of the systems with $C_s = 2.2 \times 10^{-3}$ M, and $\Phi = 0.1, 0.2$ and $0.3$.

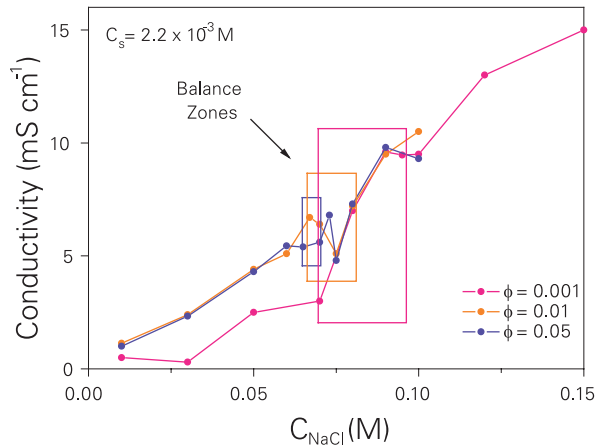

Figure 8: Conductivity of the systems with $C_s = 2.2 \times 10^{-3}$ M, and $\Phi = 0.001, 0.01$ and $0.05$.

Yithanllili Bastidas, Lisset Hernández, Issarly Rivas, Kareem Rahn-Chique and German Urbina-Villalba

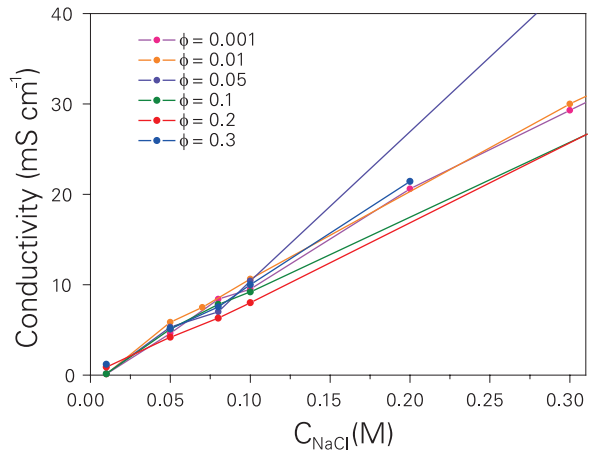

Figure 9: Conductivity of the systems with $C_s = 4.3 \times 10^{-4}$ M, and $\Phi$ = 0.001, 0.01, 0.05, 0.1, 0.2, and 0.3.

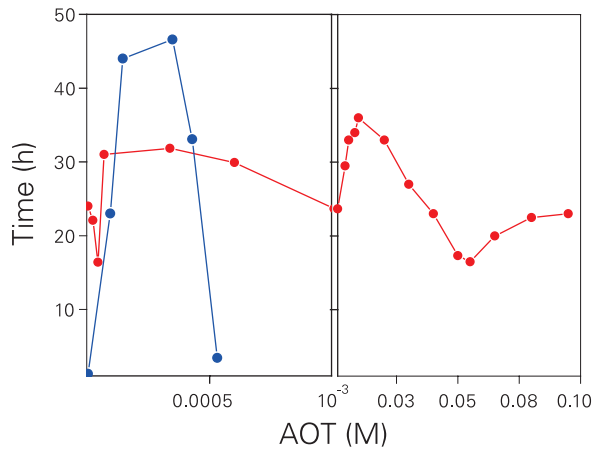

Figure 10: Stability of the emulsions with respect to the surfactant concentration ($\Phi$ = 0.0083). Set I ($C_{NaCl}$ = 0.01 M) red symbols, Set II ($C_{NaCl}$ = 0.05) blue circles.

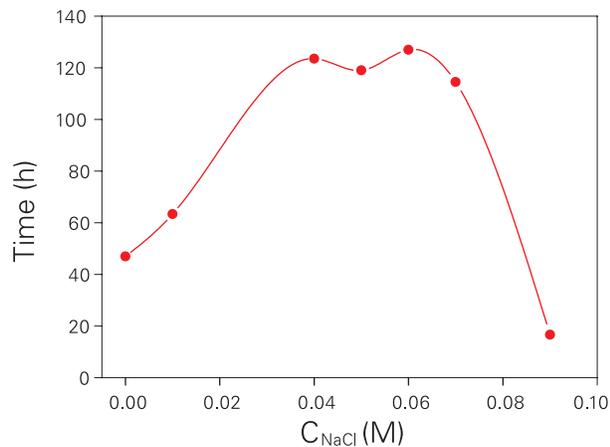

Figure 11: Stability of the emulsions with respect to the salt concentration for $C_s = 2.2 \times 10^{-3}$ M, and $\Phi$ = 0.05.

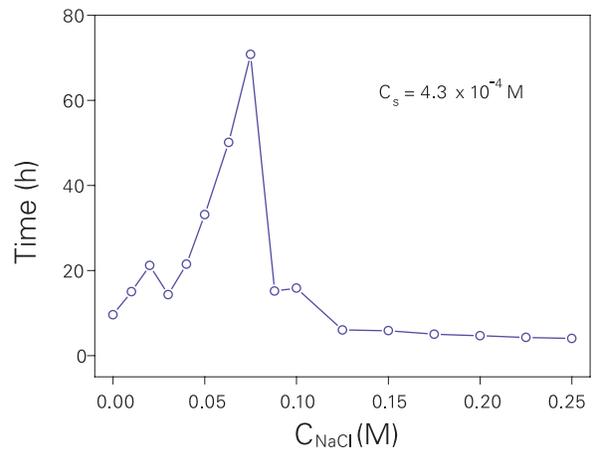

Figure 12: Stability of the emulsions with respect to the salt concentration for $C_s = 4.3 \times 10^{-4}$ M, and $\Phi$ = 0.0083.

surfactant adsorbed. As a recent article shows [Urbina-Villalba, 2013], the electrostatic potential of a dodecane-in-water nanoemulsion stabilized with sodium dodecyl sulphate, shows a maximum of stability as a function of the ionic strength. This happens around a surfactant concentration of 40 mM.

According to Fig. 10, the stability of the decane/AOT/water system at low volume fraction of oil ($\Phi$ = 0.0083) shows a secondary maximum at 0.34 mM, an absolute maximum at 9.5 mM, and secondary minima at $5 \times 10^{-5}$ M, 1 mM, and 55 mM. A similar behavior was obtained using a transmittance of 60% as a stability criteria, instead of the value of 50% previously employed. The origin of this behavior cannot be explained. The measurements of Set II were carried out by a different operator with a different set of solutions but using a salinity of 0.05 M. Again, they confirmed the existence of a maximum of stability at very low surfactant concentrations (0.01 mM < $C_s$ < 0.5 mM).

Figures 11 and 12 illustrate the dependence of the stability of the emulsions with respect to the salt concentration. In both figures the systems show a maximum of stability very close to the balance zone. In the former case, the maximum value occurs at $C_{NaCl}$ = 0.065 M. In the latter, it occurs at 0.075 M. The data of Fig. 11 can be related to the phase diagram of Fig. 3 and the conductivity graphs of Figs. 7 and 8. As the volume fraction of oil diminishes, the balance zone moves toward higher salt concentrations. Hence it appears that the maximum of stability shown in Fig. 11 is very close but probably outside the balance zone.

The situation is even more complex for Fig. 12, since the lowest volume fraction reported in the diagram of $C_s = 4.2 \times 10^{-4}$ M (Fig. 4) is 0.001. Hence, the pronounced decrease of stability observed in Fig. 12 for $C_{NaCl} > 0.075$ M, apparently corresponds to the $(\underline{2} \rightarrow \overline{2})$ transition.

The above findings coincide with the predictions of Ruckenstein based on the Gibbs adsorption equation and thermodynamic arguments [Ruckenstein, 1997]. He concluded that maximum surfactant adsorption occurs at both sides of the balance zone. For non-ionic surfactants, a deep minimum surfactant adsorption occurs just at the phase inversion temperature (PIT). Hence, minimum stability is expected at the PIT, and maximum stability is expected in the adjacent regions. This coincides with our findings except for the fact that we did not observe pronounced region of minimum stability. Under the current experimental conditions, it is either too narrow to be detected with our formulation scan, or non-existent.